\documentclass[
  reprint,
  superscriptaddress,
  amsmath,
  amssymb,
  aps,
  prb,
  floatfix,
  showkeys
]{revtex4-2}

\usepackage{graphicx}
\usepackage{dcolumn}
\usepackage{bm}
\usepackage[T2A]{fontenc}    %
\usepackage[utf8]{inputenc}              
\usepackage[russian, english]{babel}
\usepackage[colorlinks]{hyperref} 
\begin{document}
\preprint{APS/123-QED}

\title{Resonant Photoluminescence of Quantum Incompressible Liquids}

\author{D.~A.~Shchigarev}
\email{dima.shigarev@yandex.ru}
\affiliation{%
Institute of Solid State Physics named after Yu.\,A.\,Osipyan, 
Russian Academy of Sciences, Chernogolovka, Moscow District 142432, Russia
}%
\affiliation{Moscow Institute of Physics and Technology, Dolgoprudny, Moscow Region, Russia}
\author{A.~V.~Larionov}
\author{L.~V.~Kulik}
\author{E.~M.~Budanov}
\author{I.~V.~Kukushkin}
\affiliation{%
Institute of Solid State Physics named after Yu.\,A.\,Osipyan, 
Russian Academy of Sciences, Chernogolovka, Moscow District 142432, Russia
}%

\author{V.~Umansky}
\affiliation{%
Braun Center for Submicron Research, Weizmann Institute of Science, 
Rehovot 76100, Israel
}%

\date{\today}
\begin{abstract}
We investigate resonant photoluminescence arising from incompressible quantum liquids formed in two-dimensional electron systems. We demonstrate that, for excitons composed of a photoexcited electron occupying the upper spin sublevel of the zeroth Landau level and a valence-band hole, the influence of disorder potential fluctuations on optical recombination is strongly suppressed, indicating complete screening of the disorder. We identify an optical invariant quantity that is insensitive to excitation energy yet strongly dependent on the electron temperature, serving as a probe of exciton recombination in quantum liquids. Analysis of this quantity reveals that quantum-liquid formation initiates at \(\nu = 1/3\) as the electron temperature decreases, consistent with the Laughlin state. Upon further cooling, the range of filling factors exhibiting quantum-liquid behavior expands continuously from \(\nu = 1/3\) toward \(\nu = 1/2\). Transitions between distinct incompressible quantum-liquid states occur smoothly, without well-defined phase boundaries separating insulating and conducting regimes. Locally, the system retains quantum-liquid characteristics even as bulk transport measurements indicate finite conductivity. Finally, we present a phase diagram delineating the stability region of incompressible quantum liquids relative to conductive phases.
\end{abstract}

\keywords{resonant photoluminescence, two-dimensional electrons, quantum incompressible liquids}

\maketitle

\section{Introduction}
Incompressible quantum liquids (QLs) in two-dimensional (2D) electron systems under magnetic fields at fractional Landau-level filling have been a subject of intense study for over four decades~\cite{Tsui1982,Laughlin1983}. In recent years, interest in these exotic phases has surged, driven by the experimental observation of anyonic statistics in QL quasiparticles and the concurrent development of topological quantum computing schemes utilizing anyons~\cite{Kitaev2003,Sarma2015}. Anyonic statistics manifest in transport measurements~\cite{Bartolomei2020,Nakamura2020,Werkmeister2025}. Since QLs are dielectric, these measurements primarily probe the one-dimensional conducting edge channels of the 2D electron system, thereby allowing indirect access to the bulk properties of the QLs.

Optical methods are essential for directly probing the bulk properties of QLs and their associated neutral excitations~\cite{Rosenow2009,Kulik2021}. A theoretical connection between the neutral excitations of the Laughlin QL and its underlying geometrodynamics has been established~\cite{Haldane2011,Golkar2016}. This insight has led to the experimental observation of novel quasiparticles, such as chiral gravitons~\cite{Liou2019,Liang2024}, as well as their spin-1 analogs~\cite{Kulik2021}.

In this work, we present experimental results on the photoluminescence (PL) of QLs, offering new insights into their bulk properties and the nature of phase transitions between distinct QL states. Modeling the PL spectrum of QLs presents significant challenges, as demonstrated by previous studies on partially filled zeroth Landau levels in GaAs/AlGaAs quantum wells~\cite{Kulik2020}.
The PL spectrum is expected to reflect the equilibrium distribution of electrons among the spin sublevels of the zeroth Landau level. When the upper spin sublevel is depopulated (i.e., when the electron filling factor drops below 1), the corresponding PL intensity should vanish. In this regime, emission should originate solely from the lowest spin sublevel of the zeroth Landau level~\cite{Goldberg1991,Dahl1992}.

This expectation holds in low-mobility electron systems, where the radiative recombination time of photoexcited carriers is significantly longer than the relaxation time of holes in the valence band. In this case, photoexcited holes relax to the upper spin sublevel of the zeroth Landau level within the heavy-hole subband. According to the selection rules for GaAs/AlGaAs quantum wells, these holes then recombine with equilibrium electrons occupying the lowest energy spin sublevel.

However, experimental results indicate that interband optical transition probabilities in QLs are not solely determined by the equilibrium electron distribution. They are also strongly influenced by the Coulomb interaction between the photoexcited electron and hole (the excitonic effect)~\cite{Schuller2003,Byszewski2006,Yusa2001}, as well as by the asymmetry between electron-electron and electron-hole interactions~\cite{Apalkov1992}.

Furthermore, under the magnetic fields relevant for QL formation, relaxation of the photoexcited electron between Landau levels is strongly suppressed. As a result, a photoexcited electron in the first Landau level can interact with a Fermi hole left in the zeroth Landau level. This hole arises in the final state of the recombination process involving an equilibrium electron.

Consequently, interpreting the PL spectrum of QLs becomes a nontrivial task, as its spectral features exhibit a strong dependence on the excitation energy~\cite{Kulik2020}. A notable simplification of the PL spectrum is observed in heterostructures with symmetrically doped quantum wells. In such systems, the wavefunctions of conduction-band electrons and valence-band holes are nearly identical along the growth direction—a condition necessary for “hidden symmetry”~\cite{Apalkov1992}.

Additional spectral simplification can be achieved through resonant photoluminescence (RPL) techniques~\cite{Kulik2020}. In RPL experiments, the excitation energy must satisfy the resonance condition \( E_L < \hbar\omega_c + E_G \), where \( \hbar\omega_c \) denotes the electron cyclotron energy in the conduction band and \( E_G \) is the energy separation between the zeroth Landau levels of the conduction and valence bands (heavy-hole subband). Under such excitation conditions, an electron is resonantly promoted from a Landau level of the valence band (either heavy- or light-hole) directly into the unoccupied upper spin sublevel of the zeroth Landau level in the conduction band.

\begin{figure}[tb]
  \centering
  \includegraphics[width=0.8\columnwidth]{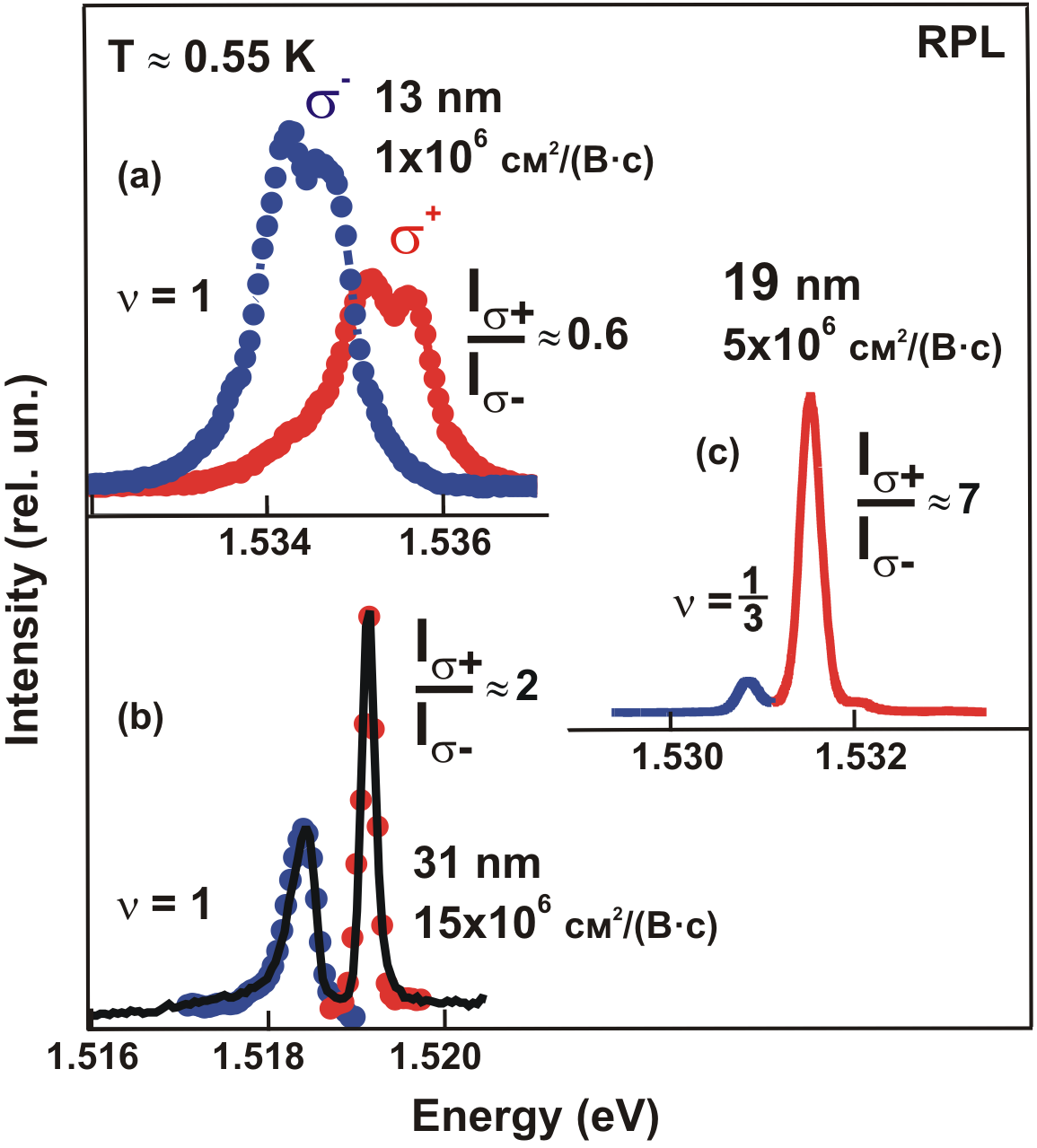}
  \caption{Polarized resonant photoluminescence (RPL) spectra of symmetrically doped GaAs/AlGaAs quantum wells recorded at a filling factor $\nu=1$ and temperature 0.55 K. Panels (a) and (b) show spectra for wells with widths of 13 nm and 31 nm, respectively, which exhibit markedly different electron mobilities. In both cases, photoexcited excitons recombine predominantly in $\sigma^-$ polarization, while equilibrium electrons recombine in $\sigma^+$. Solid lines represent the unpolarized RPL spectra; both are plotted over the same energy range. In the 13 nm well, the exciton and electron recombination lines exhibit splitting due to monolayer-scale fluctuations in well width. Panel (c) shows the RPL spectrum of the Laughlin quantum liquid at $\nu=1/3$, plotted over the same energy interval as in panels (a) and (b).}
  \label{fig:1}
\end{figure}

Such transitions occur in GaAs/AlGaAs quantum wells because the valence‑band Landau‑level hole states are not pure eigenstates of the orbital angular momentum projection along the magnetic‑field axis. Owing to heavy‑ and light‑hole subband mixing combined with strong spin–orbit coupling, these hole states form superpositions of orbital momentum projections (from \(m_\ell = 0\) to \(m_\ell = 3\)) and all spin components~\cite{D'yakonov1982,Volkov1998}.

During relaxation toward the valence‑band maximum, the photoexcited hole traverses multiple Landau levels, binding with the resonant electron to form intermediate excitonic states. If the radiative recombination time of these intermediates is shorter than the intra‑ or inter‑level hole relaxation times, additional lines emerge in the RPL spectrum. We show that the intensity ratio between these excitonic features and those originating from equilibrium‑electron recombination provides a sensitive probe of the electron system’s ground state.

\section{Experimental Details}

We investigated a symmetrically doped GaAs/AlGaAs heterostructure containing a single 19 nm quantum well. The two‑dimensional electron channel exhibited a carrier density of $0.82\times10^{11}\,\mathrm{cm}^{-2}$ and a dark mobility of $5\times10^{6}\,\mathrm{cm}^2\mathrm{V}^{-1}\mathrm{s}^{-1}$. For comparative resonant photoluminescence (RPL) measurements in both the fractional quantum Hall regime and at the integer filling factor $\nu=1$, we studied two additional symmetrically doped GaAs/AlGaAs quantum wells of widths 13 nm and 31 nm. Both wells had similar electron densities ($1.88\times10^{11}\,\mathrm{cm}^{-2}$ and $1.82\times10^{11}\,\mathrm{cm}^{-2}$, respectively) but differed markedly in mobility ($1\times10^{6}\,\mathrm{cm}^2\mathrm{V}^{-1}\mathrm{s}^{-1}$ versus $15\times10^{6}\,\mathrm{cm}^2\mathrm{V}^{-1}\mathrm{s}^{-1}$).

The samples under study were mounted in a dilution refrigerator equipped with a superconducting solenoid. Low-temperature optical measurements were conducted over the temperature range of $0.04$–$2\,\mathrm{K}$ and in magnetic fields up to $14\,\mathrm{T}$ using a two-fiber optical setup. One fiber delivered resonant excitation to the two-dimensional electron system, while the other collected the RPL signal and guided it to the entrance slit of a grating spectrometer equipped with a liquid-nitrogen-cooled CCD detector. A tunable narrow-linewidth laser with a spectral width of $10\,\mathrm{MHz}$ served as the excitation source. Polarization-resolved measurements were performed in a separate cryostat $^3\mathrm{He}$ pumped with optical windows preserving polarization.

As the quantum liquid (QL) with the largest activation gap, we selected the Laughlin QL at electron filling factor $\nu = 1/3$, the apex of the QL hierarchy~\cite{Halperin1984,Zang1993}. High-temperature resonant photoluminescence (RPL) spectra ($T > 0.5\,\mathrm{K}$) of this state were previously studied in detail in Ref.~\cite{Kulik2020}. The principal spectroscopic signature of Laughlin QL formation is a pronounced enhancement of the excitonic recombination intensity—originating from a photoexcited electron in the upper spin sublevel of the conduction-band Landau level and a photoexcited hole in the lower spin sublevel of the heavy-hole Landau subband—relative to the recombination signal from equilibrium electrons in the QL. 

The probability of radiative recombination of a single exciton is determined (up to material-dependent constants) by the coherent exciton volume, which is itself limited by random-potential fluctuations in the quantum well~\cite{Citrin1993}. Consequently, variations in the ratio of excitonic to equilibrium RPL intensities provide a measure of the spatial localization of excitonic states.

This effect is illustrated in Fig.~\ref{fig:1}, which compares the recombination intensities of photoexcited excitons and equilibrium electrons for the integer quantum Hall (QH) dielectric at $\nu = 1$ and the Laughlin QL at $\nu = 1/3$.

\section{Results and Discussion}

The case $\nu = 1$ most closely resembles the Laughlin QL at $\nu = 1/3$, as all equilibrium electrons are spin-polarized in both systems. However, at $\nu = 1$, electron–electron correlations in the conduction band do not influence the screening of the random potential, since all states in the lower spin sublevel are fully occupied. Therefore, in the integer QH dielectric, one can directly trace the effect of the random potential on the photoluminescence (PL) intensity of excitonic states.

To this end, we selected heterostructures with quantum wells differing in electron mobility by more than an order of magnitude. In the RPL spectra of these samples, two dominant lines are observed: one in $\sigma^-$ polarization and the other in $\sigma^+$ polarization (Fig.~\ref{fig:1}). These features correspond to radiative recombination between electrons with spin projections $+1/2$ and $-1/2$ in the zeroth Landau level of the conduction band and heavy holes with spin projections $-3/2$ and $+3/2$ in the valence band, respectively~\cite{Solov'ev2014}.

The higher-energy RPL line arises from excitonic recombination of a heavy hole with a photoexcited electron, while the lower-energy line originates from recombination of a heavy hole with equilibrium electrons occupying the lowest spin sublevel of the zeroth Landau level.

Increasing the electron mobility, and thus reducing the amplitude of potential fluctuations, not only narrows the RPL lines, but also leads to a three-fold increase in the ratio of photoexcited excitation to equilibrium electron intensities (Fig.~\ref{fig:1}). Thus, an increase in this ratio serves as a sensitive indicator of an enlarged coherent exciton volume. However, the enhancement observed at $\nu = 1$ is substantially smaller than that associated with the formation of the Laughlin QL (Fig.~\ref{fig:1}), despite the fact that the electron mobility in the Laughlin-QL sample is three times lower than in the high-mobility $\nu = 1$ structure.

\begin{figure}[tb]
  \centering
  \includegraphics[width=1\columnwidth]{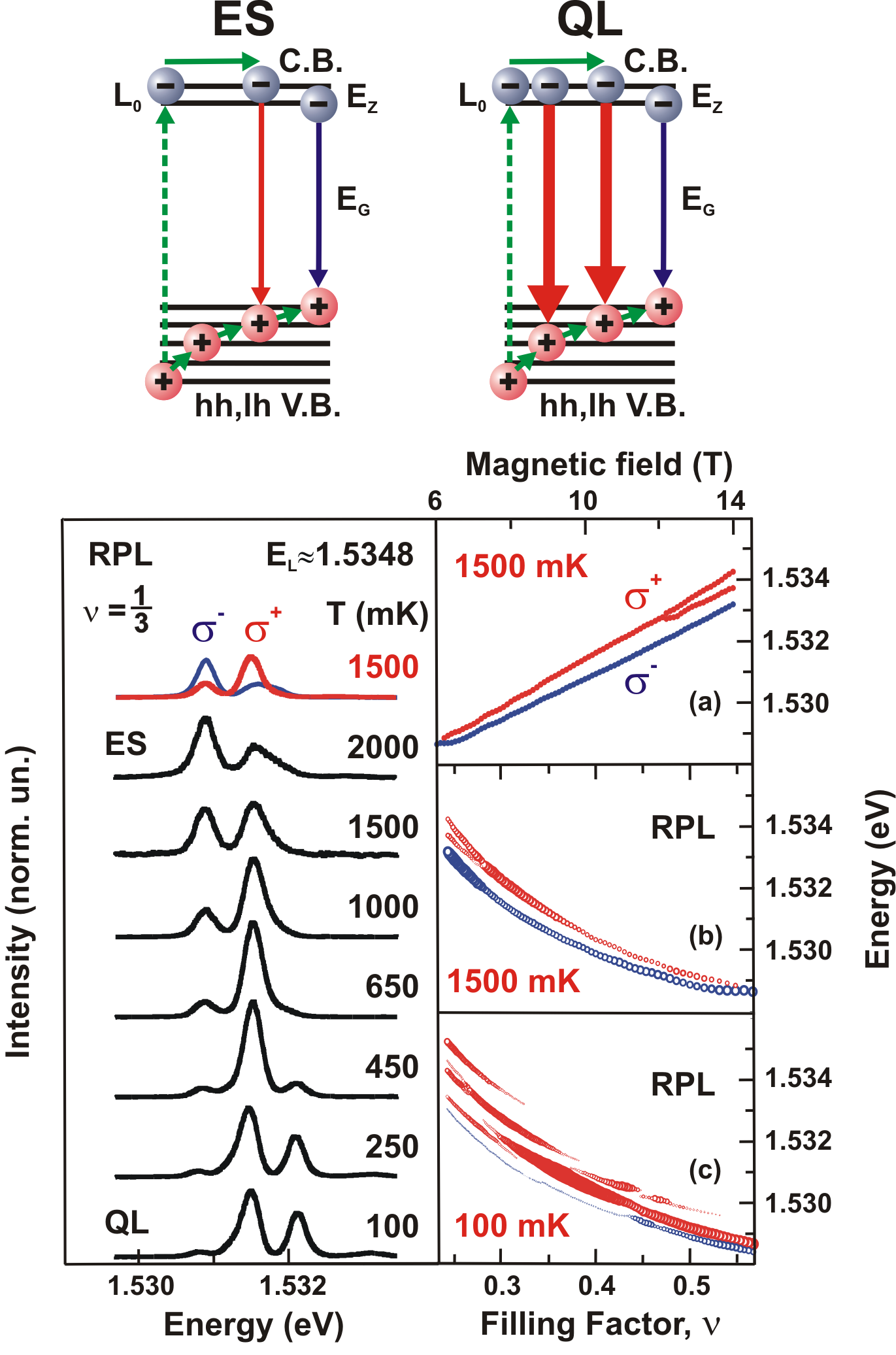}
  \caption{
  Left: Temperature evolution of RPL spectra at filling factor $\nu = 1/3$ over the range 0.10–2.00\,K under resonant excitation at photon energy 1.5348\,eV. Polarization-resolved spectra at 1.5\,K are shown at the top. \\
  Right: (a) Magnetic-field dependence of RPL peak energies measured at 1.5\,K. Red symbols correspond to excitonic recombination, and blue symbols to equilibrium-electron recombination. (b) RPL peak energies from (a) replotted as a function of filling factor; symbol sizes indicate RPL intensity. (c) RPL peak energies measured at 0.1\,K, with the same symbol conventions as in (a). The inset illustrates the optical transitions in the electron gas (ES) and in the Laughlin quantum liquid (QL).
  }
  \label{fig:2}
\end{figure}
\begin{figure}[tb]
  \centering
  \includegraphics[width=0.8\columnwidth]{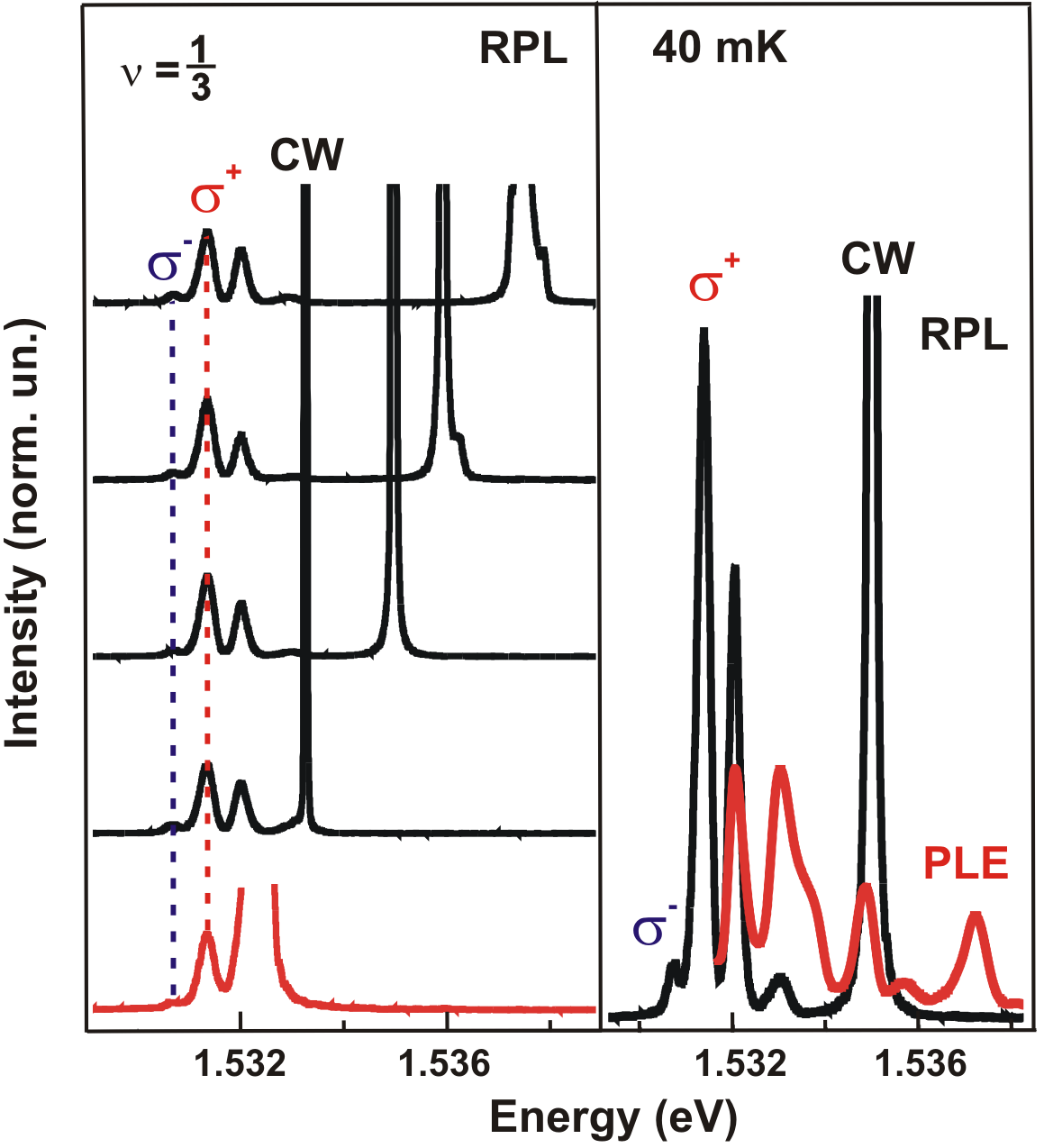}
  \caption{
  Left: RPL spectra of the Laughlin liquid at 40\,mK under varying excitation energies (black solid lines). The red line corresponds to the RPL spectrum when the excitation is tuned to one of the resonance peaks. The intensity ratio between the photoexcited-exciton ($\sigma^-$) and equilibrium-electron ($\sigma^+$) lines shows little variation with excitation energy. \\
  Right: Comparison of RPL and photoluminescence excitation (PLE) spectra, illustrating the correspondence between allowed optical transitions and RPL lines.
  }
  \label{fig:3}
\end{figure}

\begin{figure}[tb]
  \centering
  \includegraphics[width=0.55\columnwidth]{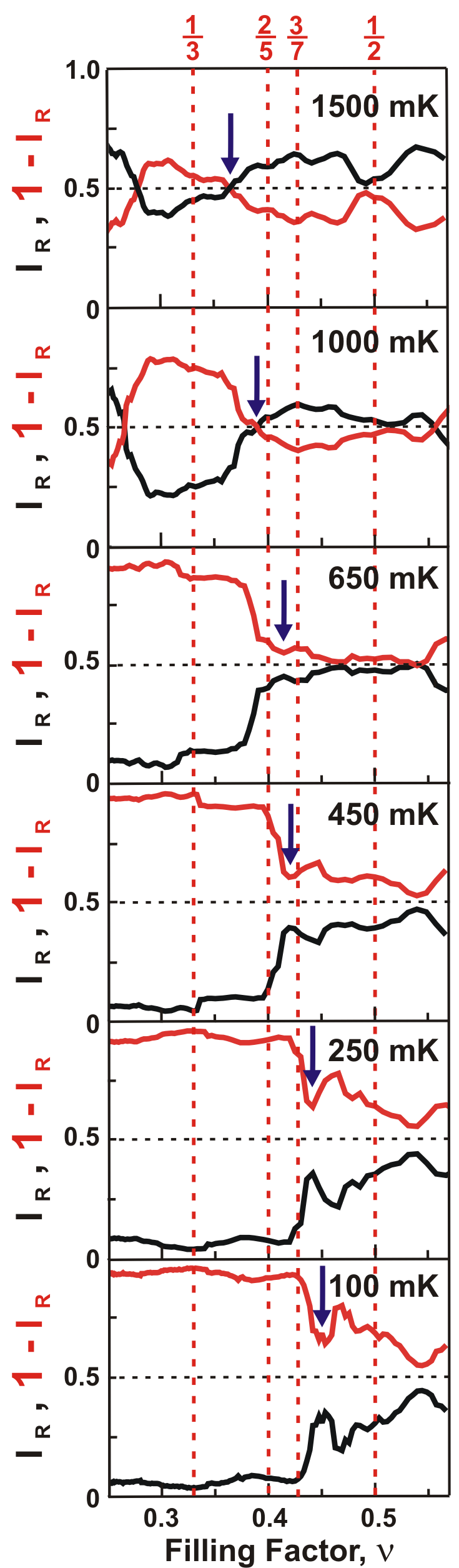}
  \caption{
  From top to bottom: variation of the free-carrier recombination to total RPL intensity ratio $(I_{R})$ as a function of filling factor (black lines) with decreasing electron temperature. For clarity, the exciton-to-RPL intensity ratio $(1–I_{R})$ is shown by red solid lines. Dashed lines indicate the filling factors corresponding to Laughlin and Jain quantum liquids. Arrows highlight kinks that delineate incompressible quantum-liquid phases from conducting phases. At 1.0 and 1.5\,K, arrows mark the intersections of the $I_{R}$ and $1–I_{R}$ curves.
  }
  \label{fig:4}
\end{figure}

To describe the ratio of RPL intensities between equilibrium electrons and photoexcited excitons at a fixed temperature ($0.55\,\mathrm{K}$), only one unknown parameter remains: the ratio of the intra-level relaxation time of photoexcited holes to the exciton recombination time. The intra-level relaxation time of photoexcited holes in the Laughlin QL was measured to be approximately $200\,\mathrm{ps}$. Direct measurement of the radiative exciton recombination time is hindered by hole relaxation processes. Considering that, in high-mobility GaAs quantum wells, the nonradiative recombination time greatly exceeds the radiative time~\cite{Kulik1997}, the exciton recombination time can be estimated from the hole relaxation time and the RPL intensity ratio. This analysis yields an exciton recombination time of approximately $30\,\mathrm{ps}$, close to the theoretical limit for a free exciton in the absence of random potential~\cite{Andreani1991}. Therefore, in the Laughlin QL, a complete exciton screening regime is realized from the random potential of the quantum well.

Cooling the Laughlin liquid results in the emergence of additional photoexcited-exciton recombination lines in the RPL spectrum, indicating that higher-energy excitons recombine within the hole-relaxation time to the zeroth Landau level (Fig.~\ref{fig:2}). The most significant experimental finding is that, at fixed temperature, the ratio of free-carrier recombination intensity to exciton recombination intensity is nearly independent of the RPL excitation resonance energy (Fig.~\ref{fig:3}). For experimental convenience, we define the $I_{R}$ ratio as the free-carrier recombination intensity divided by the total integrated RPL intensity, since the latter varies with magnetic field due to shifts of the hole levels used for excitation. The $I_{R}$ ratio thus serves as an indicator of the influence of the random potential on excitonic recombination. Because in the Laughlin QL the $I_{R}$ ratio substantially exceeds analogous values observed for conducting electron states, it can be employed to study daughter QLs in the hierarchy whose apex is the Laughlin QL at $\nu = 1/3$~\cite{Halperin1984,Zang1993}.

\begin{figure}[tb]
  \centering
  \includegraphics[width=0.9\columnwidth]{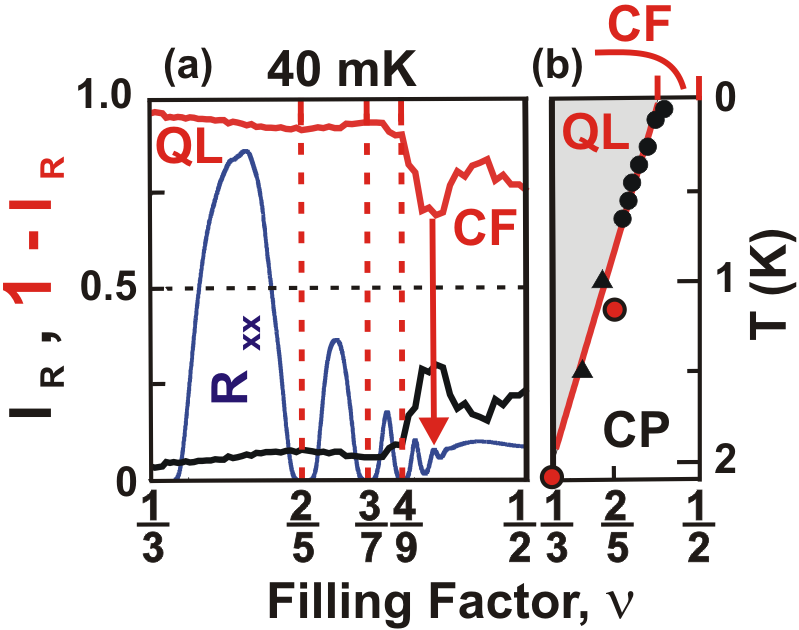}
  \caption{
  (a) $I_{R}$ (black) and $1–I_{R}$ (red) measured at $T = 40\,\mathrm{mK}$, plotted alongside the longitudinal resistance $R_{xx}$ of a similar sample at the same temperature (blue). The arrow denotes the kink separating incompressible quantum‑liquid phases from the composite‑fermion (CF) phase. \\
  (b) Phase diagram showing the filling‑factor range $\nu = 1/3$–$1/2$ occupied by incompressible quantum liquids. Black dots mark kink positions; The solid red line is a linear fit through these points and serves as the phase boundary separating the quantum liquid phase (QL) from conducting phases (CP). Red symbols are data from Ref.~\cite{Khrapai2007}, converted via electron–hole symmetry (2/3→1/3, 3/5→2/5) and the linear dependence of the QL activation gap on magnetic field.
  }
  \label{fig:5}
\end{figure}

The temperature dependence of the $I_{R}$ ratio as a function of filling factor is plotted in Fig.~\ref{fig:4}. At $T = 1.5\,\mathrm{K}$, the equilibrium‑electron recombination intensity drops below the exciton recombination intensity, marking the onset of Laughlin QL formation. Upon further cooling, the $I_{R}$ ratio decreases and the filling‑factor range over which the exciton intensity dominates broadens, continuously shifting from $\nu = 1/3$ toward $\nu = 1/2$. In regions where daughter QLs emerge, the $I_{R}$ ratio exhibits pronounced minima, indicating complete exciton screening not only in the Laughlin QL but also in its descendants. Each $I_{R}$‑versus‑$\nu$ curve shows a distinct kink separating the QL‑dominated regime (minimum $I_{R}$) from the higher‑$I_{R}$ region. Even at the lowest temperature studied ($T = 40\,\mathrm{mK}$), this kink remains below $\nu = 1/2$ (see Fig.~\ref{fig:5}), coinciding with the transport‑determined boundary between incompressible QLs and the composite‑fermion phase.

The central finding of these RPL measurements is that transitions between distinct QLs proceed continuously, without sharp phase boundaries separating bulk dielectric and conducting regimes (Fig.~\ref{fig:5}). Consequently, the electron system remains locally incompressible—even as bulk transport exhibits finite conductance. However, the $I_{R}$ ratio is not a sufficiently sensitive invariant to discriminate among different QLs; thus, RPL spectra cannot resolve transitions between Laughlin and Jain states~\cite{Laughlin1983,Jain1990} as the system moves through the hierarchy of incompressible liquids~\cite{Grigorev2022}. We hypothesize that, under conditions of finite bulk conductance, multiple QLs coexist on a local scale, whereas in fully dielectric regimes a single QL extends across the entire sample.

Based on the kink positions, we constructed a phase diagram (Fig.~\ref{fig:5}) that separates the incompressible QL regime from phases exhibiting reduced exciton screening. The onset temperatures for QL breakdown determined here are substantially lower than the activation gaps inferred from magnetotransport measurements~\cite{Du1993}. Similar QL collapse at temperatures below the transport gap was previously reported via electron recombination on neutral acceptors and capacitance spectroscopy~\cite{32,Khrapai2007}. Direct comparison with the data of Ref.~\cite{32} requires accounting for electron–hole symmetry in incompressible QLs and the linear scaling of the activation gap with magnetic field~\cite{Du1993,Khrapai2007}. After these adjustments, the results of Ref.~\cite{32} align closely with our phase boundary (Fig.~\ref{fig:5}).

\section{Conclusions}

We have studied the resonant photoluminescence (RPL) of the Laughlin liquid and related incompressible quantum liquids in the filling factor range $\nu = 1/3$ -- $1/2$. We show that, in these phases, the random‑potential fluctuations of the GaAs/AlGaAs quantum well exert a negligible effect on photoexcited excitonic states. The exciton recombination time approaches the theoretical limit for a free exciton, enabling us to track the temperature dependence of quantum‑liquid formation. Upon cooling, the Laughlin liquid at $\nu = 1/3$ appears first; further cooling does not produce a distinct Jain liquid at $\nu = 2/5$ but rather extends the incompressible regime continuously between $\nu = 1/3$ and $\nu = 1/2$. An assumption is made that, under conditions of finite bulk conductance, multiple quantum liquids coexist. We constructed a phase diagram delineating incompressible-liquid and conducting regimes. Extrapolation to $T = 0$ shows that the phase boundary remains below $\nu = 1/2$, in agreement with composite‑fermion transport data~\cite{Du1993}. Finally, we find that the collapse temperature of these quantum liquids is substantially lower than the activation gaps inferred from transport measurements~\cite{Du1993,Rosales2021}.

\begin{acknowledgments}
This work was supported by the Russian Science Foundation (Grant No.~23‑12‑00011).
\end{acknowledgments}
\nocite{*}

\bibliography{apssamp}

\end{document}